\PassOptionsToPackage{draft}{hyperref}
\documentclass[sigconf]{acmart}

\setcopyright{none}
\renewcommand\footnotetextcopyrightpermission[1]{}

\acmConference[Technical Report]{Technical Report}{2026}{}
\acmBooktitle{Technical Report}
\acmDOI{}
\acmISBN{}

\usepackage{booktabs}
\usepackage{multirow}
\usepackage{tabularx}
\usepackage{graphicx}
\usepackage{amsmath}
\usepackage{amsfonts}
\usepackage{tikz}
\usetikzlibrary{arrows.meta,positioning,calc}

\newcommand{\sys}{\textsc{GORIO}}
\newcommand{\goriogust}{G-Gust}
\newcommand{\gorioabam}{G-ABaM}
\newcommand{\goriobam}{BaM}
\newcommand{\gustspdk}{GustANN}
\newcommand{\gustgds}{GDS}
\newcommand{\motivationbox}[2]{%
  \begin{center}
  \fbox{\begin{minipage}{0.88\linewidth}
  \textbf{#1.} #2
  \end{minipage}}
  \end{center}
}

\begin{document}

\title{\sys{}: GPU-Centered Remote I/O for Graph ANNS over NVMe-oF}
\subtitle{Technical Report}

\author{Gen Zhang}
\affiliation{%
  \institution{NUDT}
  \country{China}
}

\author{Wenhao Gu}
\affiliation{%
  \institution{NUDT}
  \country{China}
}

\author{Shan Huang}
\affiliation{%
  \institution{NUDT}
  \country{China}
}

\author{Xinhai Chen}
\affiliation{%
  \institution{NUDT}
  \country{China}
}

\renewcommand{\shortauthors}{Zhang et al.}

\begin{abstract}
Graph-based approximate nearest neighbor search (ANNS) is increasingly used in
vector databases and retrieval-augmented generation services, but large vector
indexes often exceed the memory capacity of a single GPU server. NVMe over
Fabrics (NVMe-oF) provides an attractive storage-disaggregation substrate, yet
existing remote storage paths are still largely CPU-centered: the CPU forms I/O
requests, drives transport progress, and determines when GPU computation can
resume. This organization is poorly matched to graph ANNS, where the next data
access is discovered inside GPU graph traversal.

This paper presents \sys{}, a system study that extends GPU-centered local I/O
to remote storage and specializes the resulting substrate for graph ANNS over
NVMe-oF. \sys{} keeps query evolution, page-miss generation, pending-query
state, and resume decisions on the GPU, while the CPU acts only as an NVMe-oF
transport and completion proxy. The design has two layers: a GPU-direct remote
I/O path that turns local page-cache misses into split-phase remote operations,
and ANNS-specific scheduling mechanisms that overlap graph traversal with
remote page service. On a SIFT1M DiskANN-style graph workload over an RDMA
NVMe-oF path, \sys{} is
1.31$\times$ faster than the state-of-the-art remote-I/O reference path and
4.89$\times$ faster than the direct remote page-cache path.
These results demonstrate a concrete GPU-centered remote I/O substrate for
graph ANNS.
\end{abstract}

\keywords{ANNS, vector databases, GPU-centric storage, NVMe-oF, GPUDirect, disaggregated storage}

\maketitle

\section{Introduction}

Approximate nearest neighbor search (ANNS) is a core operator in vector
databases and retrieval-augmented generation (RAG) pipelines. Graph-based
indexes such as HNSW and DiskANN are widely used because they provide an
effective latency-recall tradeoff at scale~\cite{hnsw-tpami20,diskann-neurips19}.
At the same time, vector collections continue to grow beyond the memory budget
of a single GPU server, making disaggregated storage an attractive deployment
model.

NVMe over Fabrics (NVMe-oF) appears to be a natural fit for this setting:
storage capacity can be scaled independently from GPU compute, and RDMA-based
transports can provide high bandwidth and low CPU-copy overhead. However,
remote graph ANNS is not just a bandwidth problem. The search process is
fine-grained and data-dependent: each query repeatedly expands graph nodes,
computes distances, updates a candidate frontier, and discovers the next pages
to fetch. If these page misses are handled by a CPU-centered remote I/O path,
the GPU loses ownership of the search schedule exactly where the workload is
most irregular.

Two observations motivate \sys{}. \textbf{(1) Remote storage.} Local
GPU-centered storage systems such as BaM showed that GPU
threads can directly issue fine-grained storage-backed accesses through a
GPU-visible page cache~\cite{bam-asplos23}. That model is attractive for ANNS,
but production deployments often disaggregate storage from GPU compute. A
remote miss traverses queue pairs, an RNIC, a target-side NVMe stack, and a
return path before the data is visible to the GPU, so local GPU I/O mechanisms
cannot be reused without a remote completion and transport design.

\textbf{(2) ANNS specialization.} Graph ANNS does not need only GPU-accessible
data movement; it also needs page-cache residency, miss tracking, and GPU-side
pending/resume semantics for many small, data-dependent page misses. A naive
remote BaM path therefore tends to
degenerate into repeated ``GPU page miss, submit remote read, busy wait for page
ready,'' which leaves little room for useful overlap across queries.

\sys{} follows these observations with a narrower design principle: keep the
graph-search semantics on the GPU, but do not require the GPU to implement the
whole NVMe-oF transport stack. The GPU owns runnable-query selection,
outstanding page-miss state, and resume decisions. The CPU remains present only
as a thin proxy that submits NVMe-oF commands, polls completions, and publishes
completion state back to GPU-visible memory.

This paper presents this design as a working system over real NVMe-oF. To the
best of our knowledge, \sys{} is the first GPU-direct remote storage substrate
designed specifically for graph ANNS. The contribution is not simply another
backend for an existing scheduler. \sys{} extends GPU-centered local page-cache
I/O to disaggregated storage, then adds ANNS-specific scheduling and proxy
mechanisms so graph-search ownership remains on the GPU.

In summary, this paper makes the following contributions:
\begin{itemize}
    \item We identify the mismatch between CPU-centered NVMe-oF scheduling and
    GPU-resident graph ANNS traversal. To the best of our knowledge, \sys{} is the first GPU-direct remote
    storage substrate designed specifically for graph ANNS.
    \item We extend the GPU-centered local I/O model to remote NVMe-oF storage
    by converting GPU page-cache misses into split-phase remote operations with
    GPU-visible completion state.
    \item We specialize this remote I/O substrate for graph ANNS using
    persistent GPU scheduling, controlled CPU proxying, completion-ready tables,
    and lock-free request descriptors.
    \item We compare \sys{} against state-of-the-art reference paths:
    \sys{} is 1.31$\times$ faster than the remote-I/O reference path on SIFT1M
    and 4.89$\times$ faster than the direct remote page-cache path.
\end{itemize}

\section{Background and Motivation}

\subsection{GPU-Centered Local and Remote I/O}

GPU-centered storage systems move fine-grained I/O initiation closer to GPU
computation. BaM is a representative example: GPU threads access a
GPU-resident page cache, generate storage-backed page misses, and use
GPU-visible metadata to determine when a page becomes available~\cite{bam-asplos23}.
This design is a natural match for irregular GPU workloads because the
processor that discovers a miss also owns the dependent computation.

However, local GPU-centered I/O does not directly match production storage
deployments where capacity is often disaggregated from GPU compute. In an
NVMe-oF deployment, a page miss crosses an RNIC, remote queue pairs, an SPDK or
kernel transport stack, and a remote NVMe namespace before data becomes visible
to the GPU. Existing local designs do not address how GPU-generated misses are
published to a remote transport, how completions become visible to GPU code, or
how the CPU can provide transport progress without taking over application
scheduling. These issues make remote GPU-centered I/O a distinct design
problem rather than a direct reuse of local GPU storage mechanisms.

\motivationbox{Motivation 1}{GPU-centered I/O must be extended from local
devices to remote NVMe-oF storage. Production deployments often disaggregate
storage from GPU compute, so a practical GPU-centered design must preserve
GPU-owned miss generation and completion visibility across the network.}

\subsection{ANNS and GPU-Direct Storage}

Graph ANNS repeatedly expands graph nodes, evaluates vector distances, updates
candidate queues, and discovers the next pages to fetch. When the graph or
vector payload is not resident in GPU memory, page misses are created by the
graph traversal itself. This makes ANNS sensitive not only to storage bandwidth
but also to who owns pending state and resume decisions.

GPU-direct storage paths such as GDS and GPU-centered local I/O mechanisms can
move data into GPU-accessible memory, but graph ANNS also requires overlap
across many small, data-dependent misses. A conventional page-read path does
not provide page-cache residency, miss tracking, and GPU-side resume semantics.
CPU-centered ANNS systems can recover overlap by coordinating many query lanes
on the host, but then the CPU owns the irregular search schedule. Existing work
therefore does not specifically optimize the remote setting in which graph ANNS
needs both GPU-resident scheduling and NVMe-oF transport progress.

The key requirement is split-phase remote page service: GPU traversal issues or
registers a miss, the affected work becomes pending, independent graph work
continues, and the GPU resumes the dependent work when the remote page becomes
visible. \sys{} targets this requirement by extending GPU-centered local I/O to
remote NVMe-oF and specializing it for graph ANNS.

\motivationbox{Motivation 2}{Remote GPU-direct data movement is not sufficient
for graph ANNS. The system must also provide page-cache residency, miss
tracking, and GPU-side pending/resume semantics so thousands of irregular graph
page misses can overlap with useful traversal work.}

\section{Design}

\subsection{Overall Framework}

\sys{} starts from the GPU-centered local I/O model introduced by BaM: GPU
threads access a GPU-resident page cache, generate page misses, and use
GPU-visible state to decide when computation can continue. The key difference
is that \sys{} extends this model across an NVMe-oF fabric. A page miss no
longer targets a local NVMe device; it becomes a remote I/O request that must
cross RDMA queue pairs, an SPDK initiator, and a remote NVMe namespace before
the page becomes visible in GPU memory.

The design preserves the GPU-centered ownership of graph search while adding a
remote transport layer. The GPU owns query evolution, page-miss generation,
pending state, and resume decisions. CPU workers do not interpret candidate
queues, graph frontiers, or query termination. They only translate GPU-produced
page requests into NVMe-oF commands and publish completion state back to
GPU-visible memory.

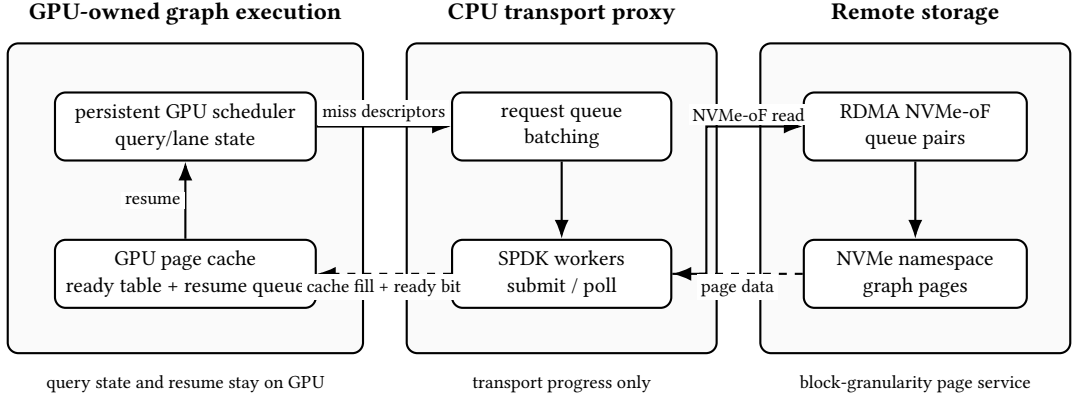
\begin{figure*}[t]
\centering
\begin{tikzpicture}[
  font=\small,
  lane/.style={draw, rounded corners, thick, fill=gray!4, minimum height=4.1cm},
  title/.style={font=\bfseries},
  block/.style={draw, rounded corners, thick, fill=white, align=center, minimum height=0.9cm},
  arr/.style={-{Latex[length=2.5mm]}, thick},
  data/.style={-{Latex[length=2.5mm]}, thick, dashed},
  note/.style={align=center, font=\footnotesize},
  lab/.style={font=\footnotesize, fill=white, inner sep=1.5pt}
]

\node[lane, minimum width=4.7cm] (gpuLane) at (0,0) {};
\node[lane, minimum width=4.1cm, right=0.55cm of gpuLane] (cpuLane) {};
\node[lane, minimum width=4.1cm, right=0.55cm of cpuLane] (remoteLane) {};

\node[title, above=0.12cm of gpuLane.north] {GPU-owned graph execution};
\node[title, above=0.12cm of cpuLane.north] {CPU transport proxy};
\node[title, above=0.12cm of remoteLane.north] {Remote storage};

\node[block, minimum width=3.45cm] (sched) at ($(gpuLane.center)+(0,0.95)$) {persistent GPU scheduler\\query/lane state};
\node[block, minimum width=3.45cm] (cache) at ($(gpuLane.center)+(0,-1.0)$) {GPU page cache\\ready table + resume queue};

\node[block, minimum width=2.9cm] (batch) at ($(cpuLane.center)+(0,0.95)$) {request queue\\batching};
\node[block, minimum width=2.9cm] (progress) at ($(cpuLane.center)+(0,-1.0)$) {SPDK workers\\submit / poll};

\node[block, minimum width=2.95cm] (nvmf) at ($(remoteLane.center)+(0,0.95)$) {RDMA NVMe-oF\\queue pairs};
\node[block, minimum width=2.95cm] (ssd) at ($(remoteLane.center)+(0,-1.0)$) {NVMe namespace\\graph pages};

\draw[arr] (sched) -- node[above,lab]{miss descriptors} (batch);
\draw[arr] (batch) -- (progress);
\draw[arr] (progress.east) -- ++(0.45,0) |- node[pos=0.72,above,lab]{NVMe-oF read} (nvmf.west);
\draw[arr] (nvmf) -- (ssd);

\draw[data] (ssd.west) -- node[below,lab]{page data} (progress.east);
\draw[data] (progress.west) -- node[below,lab]{cache fill + ready bit} (cache.east);
\draw[arr] (cache) -- node[left,lab]{resume} (sched);

\node[note, below=0.16cm of gpuLane.south] {query state and resume stay on GPU};
\node[note, below=0.16cm of cpuLane.south] {transport progress only};
\node[note, below=0.16cm of remoteLane.south] {block-granularity page service};
\end{tikzpicture}
\caption{\sys{} architecture. The GPU owns graph-search state,
page-miss generation, pending queues, and resume decisions. CPU workers are
restricted to NVMe-oF request submission, polling, and completion publication
into GPU-visible page-cache state.}
\label{fig:architecture}
\end{figure*}

\sys{} therefore contains three modules, matching Figure~\ref{fig:architecture}:
\textbf{(1) Remote storage}, which provides disaggregated NVMe capacity through
RDMA NVMe-oF; \textbf{(2) GPU-side optimization}, which maintains the page
cache, ready table, pending state, and resume queue inside GPU-visible memory;
and \textbf{(3) CPU transport proxy}, which submits NVMe-oF requests, polls
completions, and publishes page-ready state without owning graph-search
semantics.

Figure~\ref{fig:architecture} shows the \sys{} organization. GPU search
blocks execute graph traversal and access the BaM page cache. On a remote miss,
they publish request descriptors into GPU-visible request structures. CPU proxy
workers consume these descriptors, submit NVMe-oF reads through SPDK, poll
completions, and update GPU-visible completion state. The GPU scheduler then
uses this state to resume pending queries.

\subsection{Remote I/O Design}

The remote I/O layer converts a GPU page-cache miss into a split-phase NVMe-oF
operation. First, GPU code allocates a cache slot and atomically reserves a
request-descriptor slot. The descriptor records the page identifier, target
cache slot, and dependency metadata needed to resume the blocked graph-search
work, then becomes visible to the CPU proxy through a ready flag. This keeps
request allocation and dependency tracking on the GPU side instead of
serializing miss generation through a CPU-owned submit path.

Second, CPU proxy workers consume ready descriptors and submit RDMA NVMe-oF
reads through SPDK. The proxy handles transport progress but not graph-search
state. Returned page data is placed into GPU-accessible cache pages.

Third, completion publication is decoupled from query scheduling. Each GPU
page-cache slot has a GPU-visible ready-table entry. When a CPU worker completes
a remote read and fills the cache slot, it marks the corresponding ready entry.
GPU code tests this per-slot state when resuming a pending page miss, avoiding a
shared completion queue on the GPU search path.

\subsection{GPU-Side Optimization and CPU Proxy}

Graph ANNS produces many small, data-dependent misses, so the remote I/O layer
must be paired with GPU-side scheduling. \sys{} runs a persistent scheduler
kernel. Search blocks repeatedly claim work, execute graph traversal, and
either finish a query or mark it pending on a missing page. A pending query
yields the block so that the block can work on another ready query rather than
waiting in place for the remote page.

The CPU proxy is necessary because NVMe-oF/SPDK queue-pair progress is
host-managed. However, the proxy remains outside graph-search semantics.
\sys{} exposes only page-level descriptors to the CPU proxy. The proxy batches
submissions and completions, controls outstanding remote I/O, and publishes
ready state, but it does not inspect candidate queues, graph frontiers, or query
termination state. This division keeps ANNS scheduling GPU-centered while using
the CPU for the transport operations that NVMe-oF software stacks
already support efficiently.

\section{Implementation}

\sys{} is implemented in C, C++, CUDA, and SPDK. It extends a BaM-based
GPU page-cache path with an SPDK NVMe-oF initiator and integrates it with a
GustANN-derived graph-search workload. The remote target is accessed through an
RDMA NVMe-oF transport, and GPU data return uses GPU-accessible cache pages.

The implementation follows the three modules in Section~3.

\textbf{Remote storage.}
The remote storage module uses SPDK as an NVMe-oF initiator and connects to a
remote NVMe namespace over RDMA. A BaM-derived page-cache path translates a
GPU-generated graph page miss into a remote page read. The returned data is
placed in GPU-accessible cache pages, so the graph-search kernel can consume the
page without copying it through a host-side staging buffer. This module provides
capacity disaggregation while preserving the page-cache abstraction expected by
GPU-side traversal.

\textbf{GPU-side optimization.}
The GPU module contains the ANNS traversal kernels, the GPU page cache, request
descriptors, ready-table entries, and pending/resume state. When a traversal
step discovers a missing page, GPU code records the dependency and publishes a
descriptor rather than blocking in place. The persistent scheduler then switches
to other ready work. When the ready table indicates that the remote page has
arrived, the scheduler resumes the corresponding query state. This keeps
irregular graph-search control flow on the GPU.

\textbf{CPU transport proxy.}
The CPU proxy module provides the SPDK submit/poll loop needed by the NVMe-oF
software stack. Workers consume GPU-produced descriptors, submit remote reads,
poll completions, fill the corresponding GPU cache slots, and publish ready
state. The proxy also implements batching and outstanding-depth control so the
transport is served regularly without turning the CPU into a graph scheduler.
It does not inspect candidate queues, graph frontiers, or query termination
state.

\section{Evaluation}

\subsection{Setup}

\textbf{Environment.} The evaluation uses a SIFT1M DiskANN-style graph workload with
10,000 queries and recall@10 evaluation. The experiments use two servers
connected through an InfiniBand NVMe-oF path. Table~\ref{tab:platform}
summarizes the hardware and software environment. All reported times are the
benchmark search time printed by the application.

\textbf{Comparison.}
We compare five execution paths that separate the remote-I/O substrate from the
ANNS-specific scheduling layer. The short name \gustspdk{} denotes the
SPDK-backed GustANN implementation and serves
as the state-of-the-art remote-I/O reference. \goriogust{} keeps the
GustANN-style asynchronous query scheduler but replaces its remote backend with
the tuned \sys{} remote I/O path, isolating the benefit of the remote substrate.
\goriobam{} follows the BaM-style demand-paged execution model over NVMe-oF, in
which GPU page misses issue remote reads and wait for page readiness.
\gorioabam{} adds the ANNS-specific GPU scheduling and CPU-proxy mechanisms on
top of that BaM-derived remote path. \gustgds{} is a GDS-based page-read
baseline.

\textbf{Benchmark.} All systems are evaluated on the SIFT1M DiskANN-style graph workload. Each run
searches 10,000 query vectors against the same on-disk graph index and reports
top-10 nearest-neighbor results.

\textbf{Metrics.} We report end-to-end search time, throughput, and Recall@10. Throughput is
computed as $10000 / \textit{time}$ queries per second (QPS). All reported
times are the benchmark search time printed by the application.

\begin{table}[t]
\caption{Experimental platform.}
\label{tab:platform}
\centering
\small
\setlength{\tabcolsep}{4pt}
\begin{tabular}{p{1.8cm}p{5.4cm}}
\toprule
\textbf{Component} & \textbf{Configuration} \\
\midrule
Servers & Two servers \\
CPU & Intel Xeon Silver 4316, 40 CPUs per server \\
Memory & 512~GB per server \\
GPU & NVIDIA L40S, 48~GB memory \\
Network & NVIDIA ConnectX-6 InfiniBand, 200~Gbps \\
Remote SSD & Huawei HWE6AP443T8L00LN, 3.84~TB enterprise NVMe SSD \\
OS & Ubuntu 22.04 \\
CUDA & CUDA 13.0 \\
SPDK & SPDK 25.09 \\
\bottomrule
\end{tabular}
\end{table}

\subsection{End-to-End Results and Analysis}

Table~\ref{tab:progress} summarizes the end-to-end results. Each run
searches 10,000 SIFT1M queries, so throughput is computed as
$10000 / \textit{time}$.

\begin{table}[t]
\caption{End-to-end SIFT1M results over 10,000 queries.}
\label{tab:progress}
\centering
\footnotesize
\setlength{\tabcolsep}{3pt}
\begin{tabularx}{\columnwidth}{lrrX}
\toprule
\textbf{System} & \textbf{Time} & \textbf{QPS} & \textbf{Speedup} \\
\midrule
\goriogust{} & 0.99~s & 10,058 & 1.31$\times$ vs. \gustspdk{}; 121$\times$ vs. \gustgds{} \\
\gustspdk{} & 1.30~s & 7,680 & reference \\
\gorioabam{} & 1.77~s & 5,650 & 4.89$\times$ vs. \goriobam{}; 68$\times$ vs. \gustgds{} \\
\goriobam{} & 8.66~s & 1,155 & 14$\times$ vs. \gustgds{} \\
\gustgds{} & 120.19~s & 83 & baseline \\
\bottomrule
\end{tabularx}
\end{table}

The first comparison is between \goriogust{} and
\gustspdk{}. Both paths use a GustANN-style asynchronous
query scheduler, so the comparison isolates the remote-I/O substrate.
\goriogust{} improves search time from 1.30~s to 0.99~s, or
1.31$\times$ higher throughput. This indicates that the tuned \sys{} backend
provides a lower-overhead remote page service than the SPDK-backed reference
path under the same high-level scheduler. The improvement comes from extending
GPU-centered page-cache I/O to the remote setting with GPU-visible data
placement and a lightweight SPDK/NVMe-oF transport path, reducing the cost of
fine-grained graph page misses.

The second comparison is between \gorioabam{} and
\goriobam{}. These two paths share the same remote BaM-derived page-cache
substrate, so the comparison isolates the ANNS-specific optimization layer.
\gorioabam{} replaces blocking page waits with persistent GPU scheduling,
controlled transport proxying, GPU-visible completion publication, and
lock-free request descriptors. The result reduces search time from 8.66~s to
1.77~s, a 4.89$\times$ throughput improvement. This shows that the main cost in
the direct BaM-style NVMe-oF path is not simply remote bandwidth; it is the loss
of useful GPU work while graph traversal waits for remote pages. \gorioabam{}
improves performance by turning page misses into split-phase events: issue the
miss, yield blocked work, run other ready work, and resume when the page becomes
visible in the GPU cache.

The third comparison is against \gustgds{}. All
\sys{} variants substantially outperform this GDS page-read baseline:
\goriogust{} is about 121$\times$ faster,
\gorioabam{} is about 68$\times$ faster, and
\goriobam{} is about 14$\times$ faster. This confirms that
GPU-resident graph ANNS requires a fine-grained asynchronous storage path
rather than a conventional page-read path. GDS provides GPU-accessible data
movement, but it does not by itself provide the queueing, page-cache residency,
miss tracking, and GPU-side resume semantics needed to overlap thousands of
irregular graph page misses.

\motivationbox{Result Takeaway}{The results directly address the two challenges
identified in Section~2. First, \sys{} demonstrates that GPU-centered I/O can
be extended to real NVMe-oF storage and outperform the remote-I/O reference
path. Second, \sys{} demonstrates that ANNS-specific split-phase miss handling
removes the blocking behavior of a direct remote page-cache path, improving
throughput by 4.89$\times$ while preserving GPU ownership of graph search.
Together, \sys{} outperforms the GPU Direct Storage baseline by up to
121$\times$.}

\section{Related Work}

\textbf{GPU-centric storage.}
BaM showed that GPU threads can initiate
fine-grained storage-backed accesses directly through a software cache and
high-throughput queueing substrate~\cite{bam-asplos23}. \sys{} builds on this
idea by extending GPU-centered page-cache I/O from local NVMe devices to remote
NVMe-oF storage, where descriptor publication, completion visibility, and
transport proxying become first-order design issues.

\textbf{Disk-resident ANNS.}
DiskANN and Starling study page layout, graph
locality, and I/O-efficient search for disk-resident vector indexes~\cite{diskann-neurips19,starling-pacmmod24}.
\sys{} focuses on a different layer: how GPU-side graph traversal should drive
remote I/O and resume execution over an NVMe-oF fabric. To the best of our
knowledge, prior disk-resident ANNS systems do not provide a GPU-direct remote
storage path specialized for graph traversal over NVMe-oF.

\textbf{Host-centric remote storage paths.}
Kernel NVMe-oF and SPDK provide efficient remote block access, but they remain
CPU-centered in request formation and progress~\cite{nvmexpress-of11a,spdk}.
\sys{} keeps SPDK as a
transport mechanism while moving page-miss ownership and graph-search
scheduling back to the GPU.

\textbf{Asynchronous GPU-storage integration.}
Recent work on asynchronous GPU-storage access shows that split-phase
issue/consume semantics improve overlap for irregular GPU
workloads~\cite{agile-sc25}. \sys{} applies this principle to remote graph
ANNS, where fabric latency makes pending/resume behavior central.

\section{Discussion}

\subsection{Why GPU-Centered Remote I/O Matters}

The \sys{} result with a GustANN-style scheduler shows that the remote
substrate can support high-performance ANNS when paired with an effective
asynchronous scheduler. \sys{} moves this
ownership into GPU-resident execution, where graph traversal discovers page
misses and naturally owns pending/resume state.

This design has three advantages. First, query state remains local to GPU
computation, which is a clean fit for increasingly GPU-resident vector search
pipelines. Second, CPU resources are used for transport progress rather than
graph scheduling, which matters when one host serves multiple GPUs or multiple
tenants. Third, the GPU scheduler exposes page-miss and lane-state information
directly, making coalescing and selective retrieval optimizations easier to
express.

\subsection{Future Work}

Several directions can further strengthen \sys{}. First, a finer lane/page-miss
scheduler can expose more ready GPU work when only part of a query is blocked
by remote I/O. Second, miss coalescing can merge duplicate page requests across
lanes or queries before they reach the transport. Third, target-side selective
materialization can return only the graph records needed by active lanes,
reducing useful-byte amplification for page-granularity reads. Fourth,
query-aware multipath steering can distribute remote reads across multiple
NVMe-oF paths based on query state and queue pressure.

The evaluation should also be expanded beyond SIFT1M. Larger indexes and
additional datasets are needed to characterize how the remote substrate behaves
under higher graph depth, larger working sets, different recall targets, and
multi-GPU or multi-tenant deployments.

\section{Conclusion}

We presented \sys{}, a system study that extends GPU-centered local I/O to
remote NVMe-oF storage and specializes it for graph ANNS. \sys{} converts GPU
page-cache misses into split-phase remote operations, publishes completion state
back to GPU-visible memory, and keeps ANNS pending/resume decisions on the GPU
while using the CPU as a transport proxy. Evaluation shows that \sys{} is
1.31$\times$ faster than the remote-I/O reference path and 4.89$\times$ faster
than the direct remote page-cache path with the same recall. The result
demonstrates that GPU-owned
page-miss generation, pending state, and resume decisions can drive real
NVMe-oF storage without moving graph-search scheduling back to the CPU.

\nocite{pytorch-direct-arxiv21,mailthody-phd22,qureshi-phd22,helios-arxiv23,gmt-asplos24,neos-icde24,gdrec-jcrd24,reordered-pipelining-nsdi24,gids-pvldb24,tardis-apsys25,geminifs-fast25,collaborative-vector-fast25,cam-icde25,hyperion-icde25,phoenix-sc25,gustann-sigmod25,gofs-sosp25,bang-tbd25,gpcomp-tpds25,zhou-ms25,legend-arxiv26,agio-asplos26,flashanns-sigmod26,ccnvme-tos23,gnstor26,igb-thesis22}

\bibliographystyle{ACM-Reference-Format}
\bibliography{refs}

\end{document}